\documentclass[twocolumn,journal]{IEEEtran}
\usepackage[T1]{fontenc}
\usepackage{float}
\usepackage{bm}
\usepackage{amsmath}
\usepackage{graphicx}

\makeatletter

\providecommand{\tabularnewline}{\\}
\floatstyle{ruled}
\newfloat{algorithm}{tbp}{loa}
\providecommand{\algorithmname}{Algorithm}
\floatname{algorithm}{\protect\algorithmname}

\usepackage[caption=false,font=footnotesize]{subfig}
\usepackage{bm}
\usepackage{algorithm,algpseudocode}

\@ifundefined{showcaptionsetup}{}{%
 \PassOptionsToPackage{caption=false}{subfig}}
\usepackage{subfig}
\makeatother

\begin{document}

\title{Smart Grid Monitoring Using Power Line Modems: Anomaly Detection
and Localization}

\author{Federico~Passerini,~\IEEEmembership{Student Member,~IEEE,} and
Andrea M.~Tonello,~\IEEEmembership{Senior Member,~IEEE}\thanks{Federico Passerini and Andrea M. Tonello are with the Network and
Embedded System Group, University of Klagenfurt, Klagenfurt, Austria,
e-mail: \{federico.passerini, andrea.tonello\}@aau.at.}}
\maketitle
\begin{abstract}
The main subject of this paper is the sensing of network anomalies
that span from harmless impedance changes at some network termination
to more or less pronounced electrical faults, considering also cable
degradation over time. In this paper, we present how to harvest information
about such anomalies in distribution grids using high frequency signals
spanning from few kHz to several MHz. Given the wide bandwidth considered,
we rely on power line modems as network sensors. We firstly discuss
the front-end architectures needed to perform the measurement and
then introduce two algorithms to detect, classify and locate the different
kinds of network anomalies listed above. Simulation results are finally
presented. They validate the concept of sensing in smart grids using
power line modems and show the efficiency of the proposed algorithms.\end{abstract}


\section{Introduction}

\IEEEPARstart{I}{n} recent years, relevant research has been carried
out to analyze what information can be harvested about a power line
network (PLN) from the analysis of power line communication (PLC)
signals, using both the Narrowband (3-500 kHz) and Broadband (2-30
MHz) frequency ranges. The aim is to use power line modems (PLMs)
not only as mere communication devices, but also as active sensors
that can continuously monitor the status of the grid. This role is
classically absolved by phasor measurement units (PMUs) and different
kind of specific sensors displaced around the network, which work
at frequencies up to few kHz and need separate communication devices
to share information \cite{7961200}. Sensing with PLM includes sensing
and communication in a single device and enables the exploitation
of frequencies up to few tenths of MHz, which is beneficial in small
size networks as medium voltage or low voltage distribution networks. 

PLC sensing can be performed in two ways: using end-to-end communication
between two modems or reflectometry from a single modem. The classical
two-way-handshake used to establish a connection between two power
line modems (PLMs) has been exploited to gain information about the
topological structure of the grid \cite{erseghe2013topology,lampe2013tomography}.
The authors of these works propose to measure the time-of-flight of
the handshake between all the modems present in the network to estimate
the length of the connecting wiring, and use different algorithms
to infer the network topology. The frequency response of a point-to-point
communication link can be used either to detect the presence of a
fault \cite{7778781} or to monitor the aging of the cable infrastructure
\cite{7897106,6525848}. The former work relies on a direct comparison
of the channel transfer function (CTF) before and after the fault
occurrence, while the second monitors the CTF using a machine learning
algorithm and assumes an off-line training performed before. The same
problems have been tackled using a reflectometry approach, always
by analyzing the CTF of the echo coming back to the transmitting modem.
The authors of \cite{ahmed2012topology} and \cite{io} proposed different
ways to infer the network topology, while fault detection and location
has been addressed in \cite{7476292,6954542,6497543}. While the aforementioned
works on topology identification are rather developed and close to
a possible implementation, those about anomaly detection and location
are limited either to the treatment of limited parts of the problem
or to the application to small examples. We refer here and in the
following to an anomaly as a modification of the expected behavior
of the system, i.e. the power line medium.

The main aim of this paper is to propose a framework that enables
the autonomous detection and location of network anomalies in distribution
grids thanks to the PLC technology. Nevertheless, the focus of this
paper is not on the communication technology itself, but rather on
the anomaly detection and localization algorithms that can be implemented
in a PLM. The contribution starts from the results obtained in \cite{SGSI}.
Therein, a thorough analysis has been carried out to model the effect
of electrical anomalies on the signal propagation and to show which
physical quantities can be measured for the purpose of grid monitoring.
The framework proposed in this paper can be applied in medium or low
voltage distribution networks where at least one In-Band Full Duplex
(IBFD) PLM \cite{7505646} is deployed, possibly at the central office,
in order to perform reflectometry. Other PLMs can be deployed at the
termination nodes of the PLN to perform end-to-end sensing. The first
contribution of this paper consists of establishing what the required
modem architectures and measurement techniques that are needed to
perform either reflectometric or end-to-end sensing are. We propose
to measure the input network impedance and the CTF respectively, using
standard PLC to generate test signals and PLM as sensing devices.
The second contribution consists of establishing different measurement
recurrences, based on the kind of anomaly that is sought. For example,
tracking cable aging requires less frequent sensing events than detecting
a brief fault. In this regard, we present different sensing techniques
that take into account the frequency of the sensing events. Finally,
we propose different algorithms to detect and localize anomalies.
A first algorithm is used to detect and distinguish between different
kind of anomalies, and to track their evolution over time, taking
into account the time variance that characterizes power line channels
\cite{lampe2016power}. A second algorithm, which relies on the knowledge
of the network topology, is proposed to automatically localize the
detected anomaly by analyzing the sensed trace in time domain. Different
simulation results are presented that elucidate the differences between
reflectometric and end-to-end measurements, ant that show the efficiency
of the proposed algorithms.

The rest of the paper is structured as follows. Section \ref{sec:Measurements-with-PLMs}
is dedicated to the description of the required modem architectures
and to the introduction of the proposed sensing technique. Detail
considerations about the frequency of the sensing events and the appropriate
sensing algorithms to use are given in Section \ref{sec:General-considerations}.
Section \ref{sec:Anomaly-location} presents the proposed anomaly
detection and location algorithms. Extensive simulation results are
presented in Section \ref{sec:Results} and conclusions follow in
Section \ref{sec:Conclusions}.

\section{Monitoring with PLMs\label{sec:Measurements-with-PLMs}}

In this section we summarize relevant background information and present
different measurement architectures to perform network sensing with
PLMs.

\subsection{Background}

Both reflectometric and end-to-end sensing can be used to monitor
a generic PLN. In particular, with the reflectometric approach both
the input reflection coefficient $\bm{\rho}_{\mathbf{in}}$ and the
input admittance $\mathbf{Y_{in}}$ can be measured, while end-to-end
monitoring is based on the measurement of the CTF $\mathbf{H}$. Monitoring
is performed by comparison of the present measurement with a previous
one that refers to an unperturbed state of the network. Based on the
model used to describe the effect of the anomaly, the comparison is
different: it consists of a division, if the so called chain model
is used, or a subtraction, if the superposition model is used. The
resulting trace presents peculiar characteristics that allow us to
identify the presence and the type of the anomaly. When the trace
is analyzed in time domain, it also provides information about the
location of the anomaly. As for the quantity to be measured, $\bm{\rho}_{\mathbf{in}}$
does not provide information when used in combination with the chain
model, while it is as informative as $\mathbf{Y_{in}}$ when used
in combination with the superposition model. Confronting finally the
reflectometric and end-to-end approaches, with the first approach
it is easier to localize an anomaly, while the second approach can
better detect anomalies that are far away from the receiver PLM. More
details on these outcomes and their derivation using multiconductor
transmission line theory can be found in \cite{SGSI}.

\subsection{Reflectometric sensing\label{sub:Reflectometric-measurements}}

Full duplex PLMs are needed to sense $\mathbf{Y_{in}}$ or $\bm{\rho}_{\mathbf{in}}$,
since both the transmitted and the received signal have to be monitored
at the same time. The PLM transceiver can be designed to this purpose
using different architectures, based on the environment where the
modem will be deployed. Such architectures include classical schemes
like circulators, balanced bridges and current-voltage sensors. A
recent article showed that, among the mentioned architectures, the
one that yields the best quantity-to-noise ratio (QNR) is the circulator
\cite{7994720}. In this context, the QNR refers to the ratio between
the expected value of $\mathbf{Y_{in}}$ or $\bm{\rho}_{\mathbf{in}}$
and the noise related to it, similarly to the more commonly used SNR.
\begin{figure}[tb]
\begin{centering}
\includegraphics[width=1\columnwidth]{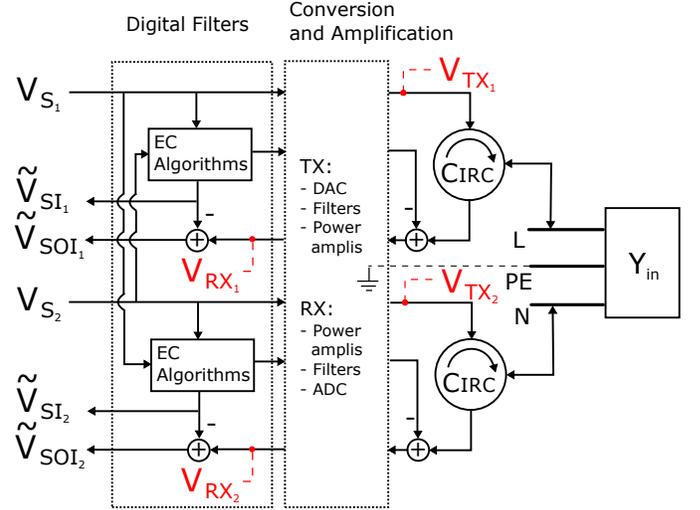}
\par\end{centering}

\caption{Proposed architecture of the full duplex PLM. \label{fig:Proposed-architecture-of}}
\end{figure}
 The circulator has also been proposed as hybrid coupler for IBFD
PLC \cite{isplc2018,7505646}. Therefore, the same modem architecture
can be used both for communication and sensing purposes.

The IBFD PLM architecture needed for the proposed system is depicted
in Fig. \ref{fig:Proposed-architecture-of}.  The system consists
of a MIMO PLM with two transmitting and two receiving channels; the
transmitting and receiving ports of each channel are connected to
a circulator, which is also connected to the power line. The digital
source signal $\mathbf{V_{S}}=\left[V_{S_{1}},V_{S_{2}}\right]^{T}$,
where the superscript $T$ denotes the transpose operation, is converted
to analog domain, thus becoming the transmitted signal $\mathbf{V_{TX}}=\mathbf{V_{S}}+\mathbf{N_{TX}}$,
where $\mathbf{N_{TX}}$ is the noise introduced by the transmission
chain. The circulators forward the signal to the PLN and the resulting
echo is forwarded to the receiver, such that the received signal $\mathbf{V_{RX}}$
is 
\begin{equation}
\mathbf{V_{RX}}=\mathbf{V_{SI}}+\mathbf{V_{SOI}}+\mathbf{N_{RX}}+\mathbf{N_{PL}}=\mathbf{V_{SI}}+\mathbf{V_{N}},\label{eq:Vrx}
\end{equation}
where $\mathbf{V_{SI}}$ is the echo signal, also called self interference,
\begin{multline}
\mathbf{V_{SI}}=-\mathbf{Y_{0}}^{-1}\bm{\rho}_{\mathbf{in}}\mathbf{Y_{0}}\mathbf{V_{TX}}\\
=-\mathbf{Y_{0}}^{-1}\bm{\rho}_{\mathbf{in}}\mathbf{Y_{0}}\mathbf{V_{S}}-\mathbf{Y_{0}}^{-1}\bm{\rho}_{\mathbf{in}}\mathbf{Y_{0}}\mathbf{N_{TX}}.\label{eq:Vsi}
\end{multline}
 $\mathbf{Y_{0}}$ is the input impedance at the channel port of the
circulator and $\bm{\rho}_{\mathbf{in}}$ is defined in \cite[Eq. 3]{SGSI}.
$\mathbf{V_{SOI}}$ is the PLC signal coming from a far-end ($\cdot_{FE}$),
also called signal-of-interest,
\begin{equation}
\mathbf{V_{SOI}}=\mathbf{H}\mathbf{V_{S_{FE}}}+\mathbf{N_{TX_{FE}}}.
\end{equation}
 $\mathbf{N_{RX}}$ is the noise introduced by the receiver stage
and $\mathbf{N_{PL}}$ is the network noise. Since $\bm{\rho}_{\mathbf{in}}$
is of interest for reflectometric sensing, it has to be estimated
based on the measurement of $\mathbf{V_{RX}}$. We remark that $\mathbf{Y_{in}}$
is directly derived from $\bm{\rho}_{\mathbf{in}}$ as
\begin{multline}
\mathbf{Y_{in}}=\left(\mathbf{I}+\bm{\rho}_{\mathbf{in}}\right)\left(\mathbf{I}-\bm{\rho}_{\mathbf{in}}\right)^{-1}\mathbf{Y_{0}}\\
=\mathbf{Y_{0}}\left(\mathbf{V_{TX}}-\mathbf{V_{SI}}\right)\left(\mathbf{V_{TX}}+\mathbf{V_{SI}}\right)^{-1},
\end{multline}
where $\mathbf{I}$ is the identity matrix. 

A series of so called Echo Cancellation (EC) techniques can be used
either in the analog \cite{isplc2018} or in the digital stage \cite{7505646}
of the receiver to estimate $\bm{\rho}_{\mathbf{in}}$, with different
accuracy based on the combination and type of algorithms used. We
point out that these techniques are already used in full-duplex PLMs
in the PLC context. In fact, in communications the receiver is only
interested in the signal $\mathbf{V_{SOI}}$, while the echo $\mathbf{V_{SI}}$
generates a self-interference that hinders the reception of the far-end
signal. Therefore, an estimate $\mathbf{\tilde{V}_{SI}}$ of $\mathbf{V_{SI}}$
is first obtained using the EC algorithms and then it is subtracted
from $\mathbf{V_{RX}}$, so that the processing stage receives only
$\mathbf{V_{SOI}}$ plus the network and hardware noises, as in classical
half-duplex communications. In the context of network sensing, the
same structure is applied, with a slight difference: $\mathbf{\tilde{V}_{SI}}$
is not only subtracted from $\mathbf{V_{RX}}$ for the communication
purpose, but can also be further processed to detect and localize
anomalies, as discussed in Section \ref{sec:Anomaly-location}.

Since the PLC channel is intrinsically linear periodic time variant
(LPTV), with periodicity equal to half the mains cycle, classical
EC algorithms such as the Least-Mean-Squares (LMS) would yield poor
results on average. An algorithm as already been proposed that tracks
channel variations in the first mains half cycle and saves the respective
state (see \cite[Algorithm 1]{7505646}). From the second half cycle
onward, a separate LMS algorithm is run for each one of the channel
states. The algorithm is also able to track load impedance variations,
which are identified by strong mean square error (MSE) registered
both at saved or non-saved symbol indexes. In Section \ref{sec:Anomaly-location},
we present a novel algorithm that enables also the sensing of faults
and cable degradations.

\subsection{End-to-end sensing\label{sub:End-to-end-measurements}}

In order to sense $\mathbf{H}$, half-duplex PLMs can be used to acquire
$\mathbf{V_{RX}}$ and classical estimation techniques can be used.
When the transmitted signal is known, an LMS algorithm can be used
to estimate $\mathbf{H}$ the same way as presented in the previous
section for $\mathbf{V_{SI}}$ and $\bm{\rho}_{\mathbf{in}}$. However,
transmitting known signals results in no communication between the
two ends. If we want to maintain a communication link between the
two modems, pilot-based or even blind CTF estimators \cite{4408690}
have to be used to estimate $\mathbf{H}$. The type of the transmitted
signals has therefore to be chosen in order to keep a balance between
high-rate communications and CTF estimation accuracy. PLC standards
\cite{nbstd,bbstd} already include a number of pilot carriers in
the OFDM symbols that are used to perform channel estimation. Pilots
can be arranged in different ways: full pilot carriers at regular
intervals (block-type), constant carrier indexes over time (comb-type)
or index swapping in consecutive symbols. Block type systems have
better convergence than the others in linear time invariant systems
(LTI), while in the case of LPTV systems like PLC channels, comb-type
or index swapping systems have been shown to yield better performance
(see \cite{6868858,1033876} and references therein). 

We finally remark that an algorithm similar to \cite[Algorithm 1]{7505646}
can be applied to the estimation of $\mathbf{H}$ and it can be used
to track periodic variations of the channel.

\subsection{Considerations on monitoring methods\label{sec:General-considerations}}

\begin{table*}[tb]
\centering{}\caption{Technical solutions, advantages and limitations of sensing with PLMs.\label{tab:Technological-solutions-and}}
\begin{tabular}{|c|c|c|}
\cline{2-3} 
\multicolumn{1}{c|}{} & End-to-end & Reflectometry\tabularnewline
\hline 
Solutions & Pilot-based estimation techniques & Adaptive filters based on completely known $\mathbf{V_{S}}$\tabularnewline
\hline 
Advantages & Use of common half-duplex modems & Sensing at symbol level preserving full data rate \tabularnewline
\hline 
Limitations & Sensing to the detriment of data rate & Use of more complex full-duplex modems\tabularnewline
\cline{1-1} 
\multicolumn{1}{c|}{} & Not continuous monitoring with block-type pilots & Presence of $\mathbf{V_{SOI}}$ might hinder estimation accuracy\tabularnewline
\multicolumn{1}{c|}{} & Not completely known $\mathbf{V_{S_{FE}}}$ with comb-type pilots & Statistics of $\mathbf{V_{S}}$ influences the estimation error\tabularnewline
\cline{2-3} 
\end{tabular}
\end{table*}
 When monitoring a network, particular attention has to be payed both
to the sensing signals used and the periodicity of the sensing events.
Regarding the sensing signals, they influence the accuracy of the
estimation of $\mathbf{H}$, $\mathbf{Y_{in}}$ or $\bm{\rho}_{\mathbf{in}}$
in different ways. In the reflectometric case the sensing signal is
known, but its statistics influences the quality of the estimation
\cite{Haykin:1996:AFT:230061}. In the case of end-to-end sensing,
the sensing signals are represented by the pilot symbols used in communication
protocols. The use of pilots intrinsically yields lower performance
than knowing the sensing signal at each subcarrier, as in the reflectometric
case. Hence, a lower performance in the estimation of $\mathbf{H}$
w.r.t. $\mathbf{Y_{in}}$ and $\bm{\rho}_{\mathbf{in}}$ is in general
expected. Technical solutions and limitations for the reflectometric
and end-to-end sensing approaches are summarized in Table \ref{tab:Technological-solutions-and}.

Regarding the periodicity of the sensing events, it depends on the
convergence time of the estimation methods used and is in general
a multiple of the symbol rate. The duration of an OFDM symbol in PLC
is in the order of hundreds of microseconds, while the effect of the
shortest anomalous events, like arching faults, lasts for some tenths
of milliseconds. This means that a convergence time of tenths to hundreds
of symbols is enough to capture the shortest anomalies. The status
of the PLN at high frequency actually varies as often as a couple
of OFDM symbols (\textasciitilde{}1 ms) due to the LPTV behavior of
the channel. All these variations are tracked as explained in Section
\ref{sub:Reflectometric-measurements} and are not considered as anomalies,
since they belong to the normal operation of the network.

Although using every OFDM symbol for sensing enables high resolution,
it is actually needed only to sense anomalies that have no permanent
effect on the system but can still be a threat, like lightning strikes,
arching faults, animal or tree temporary contact with the line and
others. On the other end, the anomalies that cause permanent or lasting
damage to the network do not require to be sensed with such rate.
In this case, $\mathbf{H}$, $\mathbf{Y_{in}}$ or $\bm{\rho}_{\mathbf{in}}$
can be estimated at time intervals that are considerably greater than
the length of a communication symbol. In particular, since typical
PLC systems are aware of the mains cycle period \cite{handbookplc},
we propose the following: 
\begin{itemize}
\item sensing at symbol level is performed using the techniques presented
above in order to identify temporary anomalies. We call this symbol
level sensing (SLS). 
\item at intervals $T$ that are multiples of half the mains period, the
channel is sensed using known values of both $\mathbf{V_{S}}$ and
$\mathbf{V_{S_{FE}}}$ . We call this mains level sensing (MLS).
\end{itemize}
We remark that the estimation techniques used for the SLS are anyhow
used anytime a PLM wants to communicate with other modems, so the
overload generated by sensing is just due to the anomaly detection
and location algorithm presented in Section \ref{sec:Anomaly-location}.
Since both the end-to-end and the reflectometric SLS can track the
periodic channel variations, the unperturbed situation is also periodic
time-varying. Hence, the anomaly detection algorithm is run on every
new sensing instance with respect to the unperturbed measurement relative
to the specific sensing instant. 

As for the second sensing approach, it has two main advantages: first,
by sensing every $T$ mains cycles, we elude the time variations of
the channel and the resulting system can be considered LTI. Second,
using known signals allows us to reduce the estimation techniques
to simple averaging, which would yield over a significant amount of
samples toward null estimation error \cite{kay1993stat}. In fact,
all the noise sources, including $\mathbf{N_{RX}}$, $\mathbf{N_{TX}}$,
$\mathbf{V_{SOI}}$ and $\mathbf{N_{PL}}$, can be considered with
good approximation to have mean zero.

\section{Anomaly detection and location \label{sec:Anomaly-location}}

In this Section, we present an algorithm that can be used to both
detect and locate anomalies, as well as distinguish between localized
faults, load impedance changes and distributed faults.

\subsection{Anomaly detection and classification\label{sub:Anomaly-detection-and}}

The unperturbed situation is considered to be when, after the startup,
the estimation algorithms presented in the previous section converge
to a minimum estimation error. It has been shown in \cite{7994720},
that the noise related to the estimation of $\bm{\rho}_{\mathbf{in}}$
and $\mathbf{Y_{in}}$is zero mean if $\mathbf{V_{N}}<\mathbf{V_{S}}/10$,
which is normally the case in PLN. When using adaptive algorithms,
subspace or interpolation techniques with finite impulse response
filters in presence of noise, the MSE is always lower bounded and
positive. On the other side, the MSE tends to zero when averaging
over a large sample set. 

In the following, we consider the case of SLS algorithms with fixed
and finite parameters, such that the MSE converges to a minimum MSE$_{\infty}$.
For every new estimation step $m$, we compute the quantity 
\begin{equation}
\Delta_{sup}\left(m,n\right)=\mathbf{\tilde{A}}\left(m,n\right)-\mathbf{\tilde{A}}_{ref}\left(m,n\right)\label{eq:Deltasup}
\end{equation}
or 
\begin{equation}
\Delta_{ch}\left(m,n\right)=\mathbf{\tilde{A}}\left(m,n\right)\mathbf{\tilde{A}}_{ref}\left(m,n\right)^{-1}\label{eq:Deltachain}
\end{equation}
depending on weather the chain or the superposition model for the
anomaly has been chosen \cite{SGSI}, where $\mathbf{\tilde{A}}$
stands for $\mathbf{Y_{in}}$, $\bm{\rho}_{\mathbf{in}}$ or $\mathbf{H}$,
and $\mathbf{\tilde{A}}_{ref}\left(m,n\right)$ is a reference value
for the unperturbed situation, chosen as a mean of the previous estimated
values. If at least for one of the $n$ indexes the value of $\left|\Delta_{sup}\right|,$$\left|\Delta_{ch}\right|$
is greater than a fixed threshold (we use three times the standard
deviation of $\mathbf{\tilde{A}}_{ref}\left(m,n\right)$), then the
index $n_{max}$ of the maximum of \eqref{eq:Deltasup}, \eqref{eq:Deltachain}
is saved. This is because the value thus found might be caused by
impulsive noise, which is common in PLNs. However, if this value is
caused by an anomaly, in the following iterations a similar value
will appear at $n_{max}$. In order to reduce false positives, few
successive realizations of the increment against the same reference
will be tested. If the value of \eqref{eq:Deltasup}, \eqref{eq:Deltachain}
at $n_{max}$ is always greater than the threshold, then an anomaly
is detected. Otherwise, $\mathbf{\tilde{A}}\left(m,n\right)$ is used
to update $\mathbf{\tilde{A}}_{ref}\left(m,n\right)$.

When an anomaly is detected, $\partial_{sup}(m,t)$ or $\partial_{ch}(m,t)$
are computed as the inverse Fourier transforms of \eqref{eq:Deltasup}
and \eqref{eq:Deltachain} respectively. Their peaks are detected,
either by classical peak-detection or super-resolution techniques
as presented in Section \ref{sub:Spectral-analysis}. The first peak
of $\partial_{sup}^{Y}(m,t)$ tells already the distance of the anomaly
from the receiving modem, while an ambiguity remains in the case of
$\partial_{sup}^{H}(m,t)$. Regarding the type of the fault, a first
distinction between localized and distributed anomalies is made. As
shown in \cite[Fig. 5,6]{SGSI}, a distributed anomaly, conversely
from localized anomalies, causes a shift in the peaks in frequency
domain. Therefore, it is sufficient to test weather the peaks of $\mathbf{Y_{in}}$
or $\mathbf{H_{tot}}$ are also present in $\mathbf{Y_{in_{a}}}$
or $\mathbf{H_{tot_{a}}}$ respectively to understand if the anomaly
is localized or distributed. In the first case, the peak would be
identified in both traces, while in the second case the response will
be negative. When the anomaly is identified as localized, the time
domain trace is analyzed. If the position of the first peak of $\partial_{sup}(m,t)$
or $\partial_{ch}(m,t)$ coincides with the position of a peak of
$\mathbf{\tilde{y}_{in}}$ or $\mathbf{\tilde{h}_{tot}}$ respectively,
then the anomaly can be a load variation. To confirm this hypotesis,
we look for the presence of the same peaks after the anomaly in $\mathbf{\tilde{y}_{in}}$
or $\mathbf{\tilde{h}_{tot}}$and $\mathbf{\tilde{y}_{in_{a}}}$ or
$\mathbf{\tilde{h}_{tot_{a}}}$. In fact, if the anomaly is a load
variation, no new peaks are created in the time domain response. If
this is the case, the anomaly is identified as an impedance variation,
otherwise it is a fault. The detection and classification technique
is summarized in Algorithm \ref{alg:Anomaly-detection-and}.

\begin{algorithm}[tb]
\caption{Anomaly detection and classification algorithm\label{alg:Anomaly-detection-and}}

\begin{algorithmic}[1] 
\Require{$\mathbf{\tilde{Y}_{in_{a}}}$, $\mathbf{\tilde{Y}_{in}}$} 
\Ensure{Presence of an anomaly, Type of the anomaly} 
\While{$\max_n\Delta_{sup}^Y < \text{thr1}$}
\State{update $\mathbf{\tilde{A}}_{ref}$}
\State{transmit a new OFDM symbol}
\State{compute $\max_n\Delta_{sup}^Y$}
\EndWhile
\State{an anomaly has been detected}
\State{compute $\partial_{sup}^Y$, $\mathbf{\tilde{y}_{in_a}}$ and $\mathbf{\tilde{y}_{in}}$}
\State{compute \emph{peaks}[$\partial_{sup}^Y$], \emph{peaks}[$\mathbf{\tilde{y}_{in_a}}$] and \emph{peaks}[$\mathbf{\tilde{y}_{in}}$]}
\State{compute \emph{peaks}[$\mathbf{\tilde{Y}_{in_{a}}}$] and \emph{peaks}[$\mathbf{\tilde{Y}_{in}}$]}
\If{max\{\emph{peaks}[$\mathbf{\tilde{Y}_{in}}$]- \emph{peaks}[$\mathbf{\tilde{Y}_{in_a}}$]\} > thr2}
\State{the anomaly is a \textbf{distributed fault}}
\ElsIf{min\{\emph{peaks}[$\partial_{sup}^Y$](1) - \emph{peaks}[$\mathbf{\tilde{y}_{in}}$]\} < thr3}
\If{max\{\emph{peaks}[$\mathbf{\tilde{y}_{in}}$] - \emph{peaks}[$\mathbf{\tilde{y}_{in_a}}$]\} < thr4}
\State{the anomaly is an \textbf{impedance variation}}
\ElsIf{the anomaly is a \textbf{localized fault}}
\EndIf
\Else
\State{the anomaly is a \textbf{localized fault}}
\EndIf
\end{algorithmic}
\end{algorithm}

\subsection{Anomaly localization: one sensor}

\begin{figure}[tb]
\begin{centering}
\subfloat[]{\begin{centering}
\includegraphics[width=0.8\columnwidth]{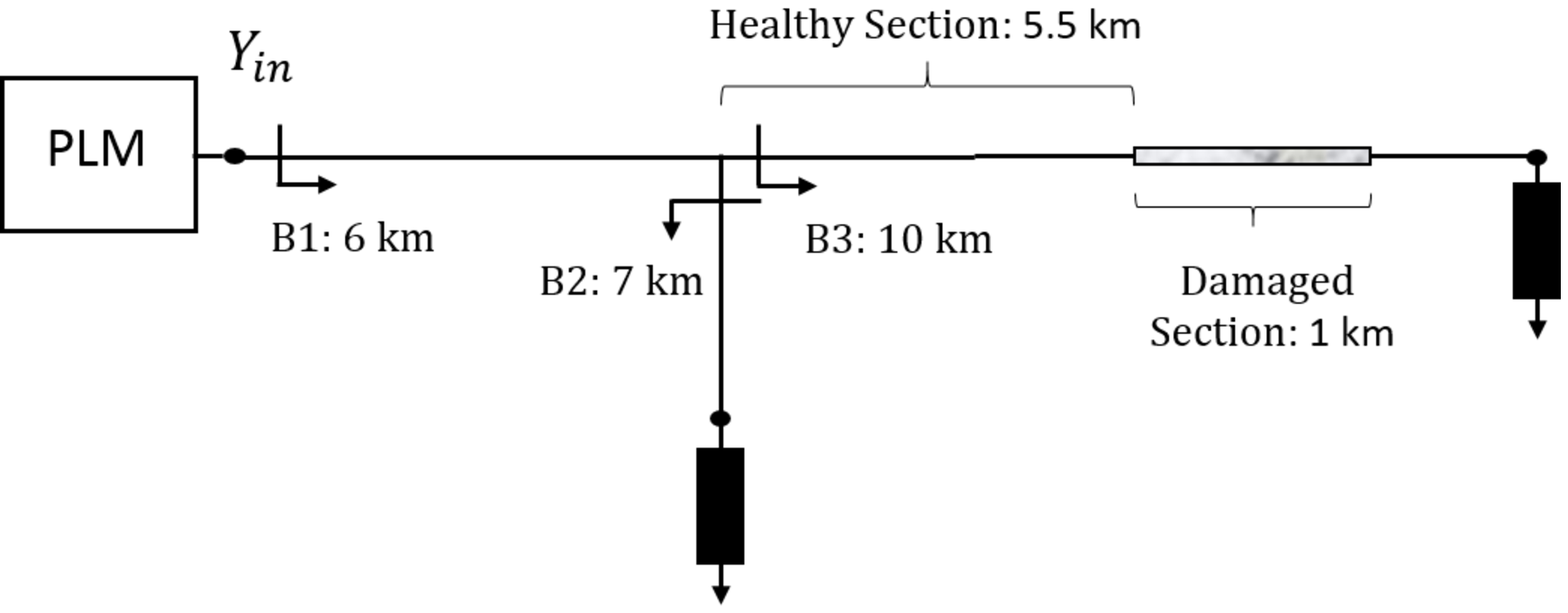}
\par\end{centering}

}
\par\end{centering}

\centering{}\subfloat[\label{fig:fig7b}]{\centering{}\includegraphics[width=0.95\columnwidth]{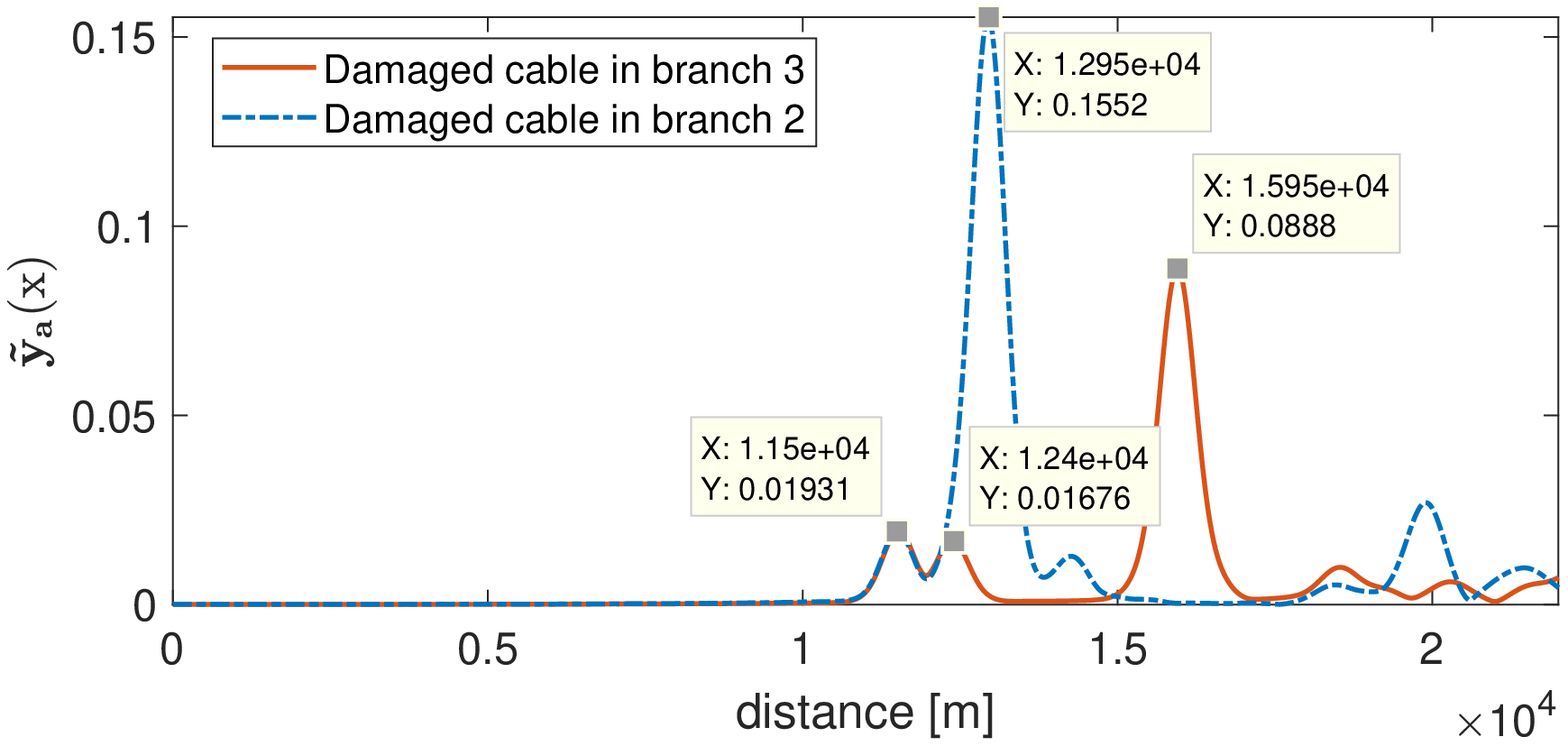}}\caption{Example of a simple network with a damaged section: a) sketch and
b) estimated admittance variation when the damaged section is on branch
2 or 3.\label{fig:Example-of-a-1}}
\end{figure}
When just one sensor or one sensor pair (in the case of end-to-end
sensing) is available, the distance of an anomaly from the measurement
point can be retrieved from the position of the first peak of $\mathbf{\tilde{y}_{a}}$
or $\mathbf{\tilde{h}_{a}}$. When the topology of the network is
known, the relative position of the peaks of $\mathbf{\tilde{y}_{a}}$
and $\mathbf{\tilde{h}_{a}}$ univocally relates every anomaly to
a precise point in the network, enabling the localization of the anomaly.
Consider the example of Fig. \ref{fig:Example-of-a-1}, where a network
with a damaged cable section 11.5 km away from the sensing point is
considered. The damaged section could be either on branch B2 or B3,
but we depicted only the B3 case for simplicity. Fig. \ref{fig:fig7b}
shows that, if the damaged section is in branch B3, then a peak appears
at 12.4 km, corresponding to the end of the damaged section and another
peak appears at 15.95 km, corresponding to the position of the load
at the end of the branch. On the other side, if the damaged section
is in branch B2, we see a prominent peak at 12.95 km corresponding
to the position of the load at the end of branch B2. This peak is
actually so high that it hides the peak at 12.4 km generated by the
end of the damaged section. Besides, we notice that the peaks corresponding
to the network loads are located nearer to the sensing point as expected
in an unperturbed situation. As explained in Section \ref{sec:Anomaly-location},
this is a clear sign that the detected anomaly is a distributed fault.
This example shows that, in the reflectometric case, it is sufficient
to analyze few peaks after the first to understand in which branch
an anomaly is located. The end-to-end sensing case is more complex,
as explained in \cite{SGSI}. In fact, the first peak of $\mathbf{\tilde{h}_{a}}$
cannot tell weather the corresponding distance is from the transmitter
or the receiver; the first peak caused by a branch ending might refer
to the ending towards the receiver or the transmitter, while in the
reflectometric case it always refers to the farther branch ending.
All of this would add significant complexity to an anomaly localization
algorithm. For these reason, the anomaly localization is not treated
in this work for the case of end-to-end monitoring.

The following algorithm (see Algorithm \ref{alg:Anomaly-location-algorithm})
can be derived to automatically locate an anomaly after it has been
detected. When analyzing $\mathbf{\tilde{y}_{a}}$, its first peak
provides an estimate $\hat{d}_{a}$ of the distance of the anomaly
from the sensing point. If the anomaly is identified as a load impedance
change, then the branch is directly identified in the hypothesis that
the network is asymmetric and there are no nodes equally distant from
the sensing point, which is common in PLN. If the anomaly is identified
as a lumped fault or a distributed anomaly, then in a first step all
the $M$ possible branches where the anomaly can be located are identified.
The distance $\mathbf{d}$ between all the nodes and the receiver
is also computed. Subsequently, for each of the $M$ possible branches,
the difference $b$ between $\hat{d}_{a}$ and the nodes $N_{1}^{m}$
and $N_{2}^{m}$ at the extremities of the $m$th branch is computed.
This step allows to identify the distance of the first few peaks after
$\hat{d}_{a}$ is the fault is in branch $m$. The result is subtracted
from $\mathbf{\tilde{y}_{a}}$, to check weather the guessed peaks
correspond to the measurement. Finally, the branch with the lowest
result is selected as the estimated anomaly branch. 

\begin{algorithm}[tb]
\caption{Anomaly localization algorithm\label{alg:Anomaly-location-algorithm}}

\begin{algorithmic}[1] 
\Require{\emph{peaks}[$\partial_{sup}^Y$], type, topology} 
\Ensure{Position of the anomaly}
\State{$d_{an}$ = \emph{peaks}[$\partial_{sup}^Y$](1)}
\State{Compute $\mathbf{d}$}
\If{type == impedance variation}
\State{$a$ = find\{$\mathbf{d}$ - $d_{an}$ < thr\}}
\State{the impedance variation is on node $a$}
\Else
\State{find the M branches where the anomaly might be}
\For{i = 1 to M}
\State{compute $b$ = $\lvert\hat{d}_{a} - [N_1^m,N_2^m]\rvert$}
\State{compute $c$ = min\{\emph{peaks}[$\partial_{sup}^Y$]-$\left(b+\hat{d}_{a}\right)$\}}
\EndFor
\State{the anomaly is located on the branch with lowest $c$}
\EndIf
\end{algorithmic}
\end{algorithm}

\subsection{Anomaly localization: multiple sensors}

In the case of multiple sensors, different techniques can be applied.
The simplest one is based on geometric considerations: the information
about the position of the first peak of $\mathbf{\tilde{Y}_{a}}$
or $\mathbf{\tilde{H}_{a}}$ coming from multiple sensors is fused
to select the point that has the expected distance from each sensor.
If the network is not symmetric, two sensing points are, in the case
of reflectometry, enough to univocally determine the branch where
the anomaly has occurred. In the case of end-to-end sensing, the presence
of multiple modems also removes the intrinsic ambiguity of the estimated
distance from the receiver. We point out that two-way end-to-end sensing
(i.e. a signal is transmitted from one modem to the other and a response
is immediately sent back to the first one) is not sufficient to solve
the position ambiguity; at least a third modem is needed.

Geometric considerations are reliable in the case of MLS, but when
it comes to SLS, the anomaly needs to be localized in a short time
frame, therefore all the sensors might need to perform the measurement
at the same time. In this case, there is a problem of interference
between the sensors, which can be alleviated by using sensing signals
that are orthogonal to each other \cite{5361355}.

Other approaches implementable with PLMs are based on the decomposition
of the time reversal operator (DORT) \cite{7335651}. These approaches
are specifically designed to detect and localize very weak faults
along the network, but they need a simulator with a complete topological
and electrical model of the network in order to work. The scattering
matrix of the network is measured before and after the fault. The
DORT is afterwards applied to find an optimum set of signals, whose
transmission is then simulated on the test network. The energy of
this optimum signals will focus on the position of the anomaly.

\subsection{Spectral analysis\label{sub:Spectral-analysis}}

As we explained in the previous section, locating an anomaly basically
turns into finding a series of peaks in the time domain response.
Due to band limitations in communication systems, especially in PLC,
the resolution might not be sufficient to separate close peaks or
might provide a too loose estimation of a peak position. However,
when the transfer function of a system can be represented as a sum
of weighted exponentials, subspace methods can be applied. This is
the case of $\mathbf{H}$, $\mathbf{Y_{in}}$ or $\bm{\rho}_{\mathbf{in}}$,
as explained in \cite[Eq. 17, 18, 27]{SGSI}. Such methods are used
to better detect and locate the presence of peaks, since they can
achieve super-resolution \cite{stoica2005spectral}.  In this research
work, we applied different subspace algorithms \cite{stoica2005spectral,1172124}
to the anomaly localization problem. Among all, the root-Music method
\cite{1172124} provided the best performance. However, when many
peaks have to be detected, their amplitude varies greatly, and the
signal bandwidth is sufficiently wide, we found that peak-location
algorithms can identify more peaks than subspace methods, even though
the resolution is lower. 

In the anomaly detection and localization problem it is more important
to identify a peak than to precisely localize it. Therefore, peak-location
algorithms are preferred in this paper over subspace algorithms.

\section{Results\label{sec:Results}}

In this Section, we present some results obtained by simulation that
show the performance in detecting and locating anomalies of the discussed
algorithms.
\begin{table}[tb]
\caption{Parameters used for the simulation\label{tab:Parameters-used-for}}

\centering{}%
\begin{tabular}{|c|c|}
\hline 
Parameter & Value\tabularnewline
\hline 
\hline 
Frequency & 4.3 kHz - 500 kHz span, 4.3 kHz sampling \cite{nbstd} \tabularnewline
\hline 
Network noise & According to \cite[Annex D.3]{nbstd} \tabularnewline
\hline 
Transmitter noise & -50 dBc (10 bit DAC, OFDM \cite{Berger:11})\tabularnewline
\hline 
Receiver noise & -60 dBc (12 bit ADC, OFDM \cite{Berger:11})\tabularnewline
\hline 
Transmitted power & According to \cite[Ch. 7]{nbstd}\tabularnewline
\hline 
Number of nodes & 20\tabularnewline
\hline 
Average branch length & 900 m \cite{pagani}\tabularnewline
\hline 
Load value & According to \cite[Annex D]{nbstd}\tabularnewline
\hline 
\end{tabular}
\end{table}

\subsection{Simulation Setup}

We developed an MTL PLN simulator using the equations presented in
\cite[Sec. 1]{SGSI}. Such simulator randomly displaces a given number
of nodes on a given surface and connects them taking into account
a maximum node degree (i.e. the number of branches connected to a
node). If not otherwise specified, we tune the simulator to displace
the nodes with average distance of 900 m, which mimics the average
displacement in a low voltage distribution network, or a medium voltage
underground distribution network. 

An anomaly in the form of a lumped impedance or a cable branch with
modified parameters can be inserted in any point of the network. The
simulator computes $\mathbf{Y_{in}}$, $\mathbf{Y_{in_{a}}}$,$\bm{\rho}_{\mathbf{in}}$,
$\bm{\rho}_{\mathbf{in_{a}}}$ at every node and $\mathbf{H_{tot}}$,
$\mathbf{H_{tot_{a}}}$ between every node pair. As for the PLM impedance,
we consider the optimum conditions for reflectometry and end-to-end
transmission. In the first case, $\mathbf{Y_{0}}$ is equal to $\mathbf{Y_{C}}$
of the cable to which the PLM is branched. In the second case, the
output impedance of the transmitter is fixed to 1$\Omega$ and the
input impedance of the receiver is fixed to 100 k$\Omega$ as typical
in half-duplex PLMs. 

As for the noise and signal powers, we use standard levels for PLC
as specified in \cite{7994720} if not otherwise stated. We finally
assume the noise introduced by the PLM coupler (hybrid and transformer)
to be negligible with respect to the other noise sources. 

Finally, in the following we do not make use of a specific EC algorithm
in the reflectometric case or a specific interpolation filter in the
end-to-end case, but we rather model the effect of the overall error
when estimating $\bm{\rho}_{\mathbf{in}}$ or $\mathbf{H_{tot}}$
on the performance of the anomaly detection and location algorithms.

\subsection{Comparison of models and measurement types}

\begin{figure}[tb]
\begin{centering}
\subfloat[\label{fig:-when-a}]{\begin{centering}
\includegraphics[width=0.43\columnwidth]{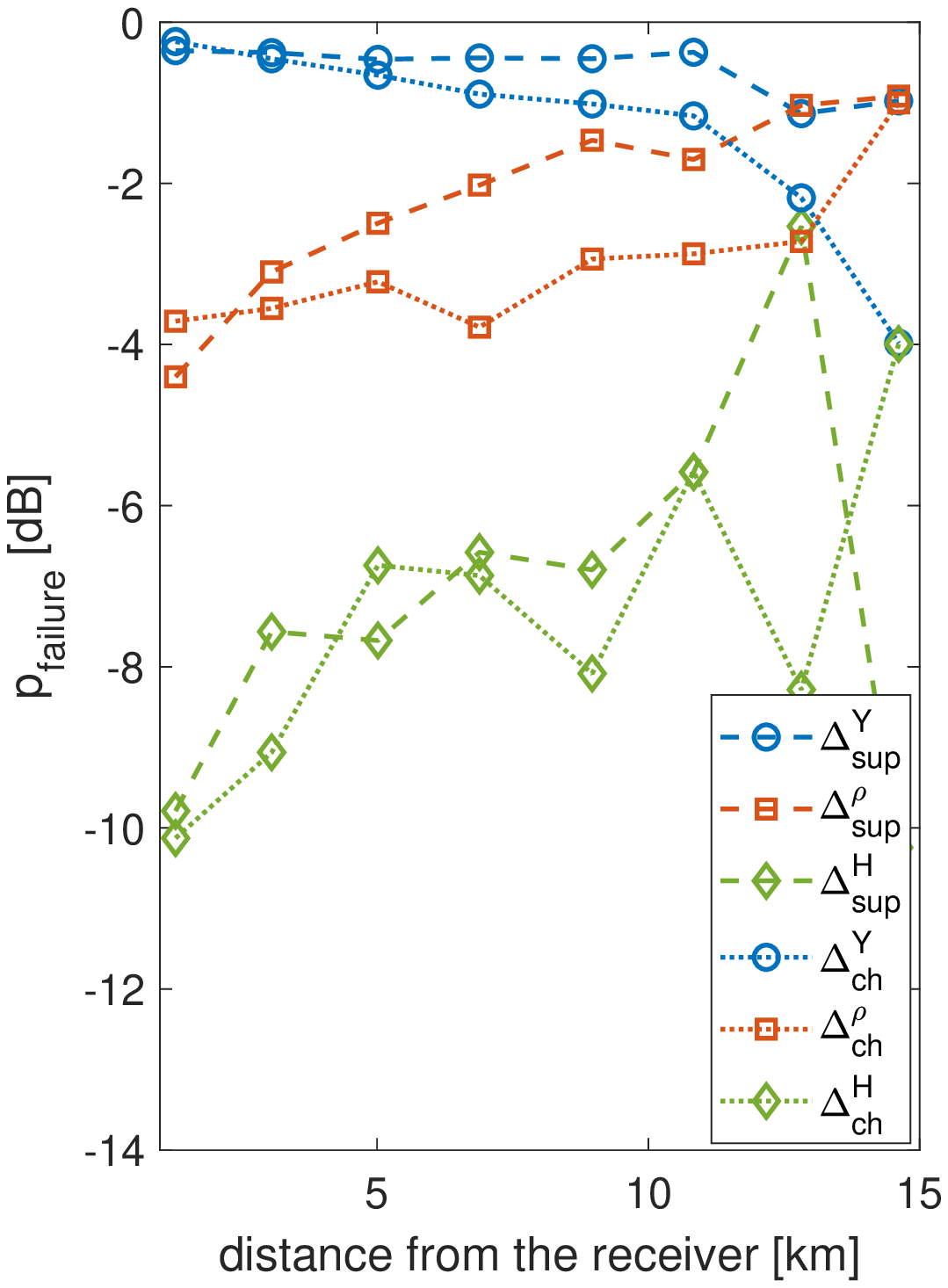}
\par\end{centering}

}\subfloat[\label{fig:-when-b}]{\begin{centering}
\includegraphics[width=0.43\columnwidth]{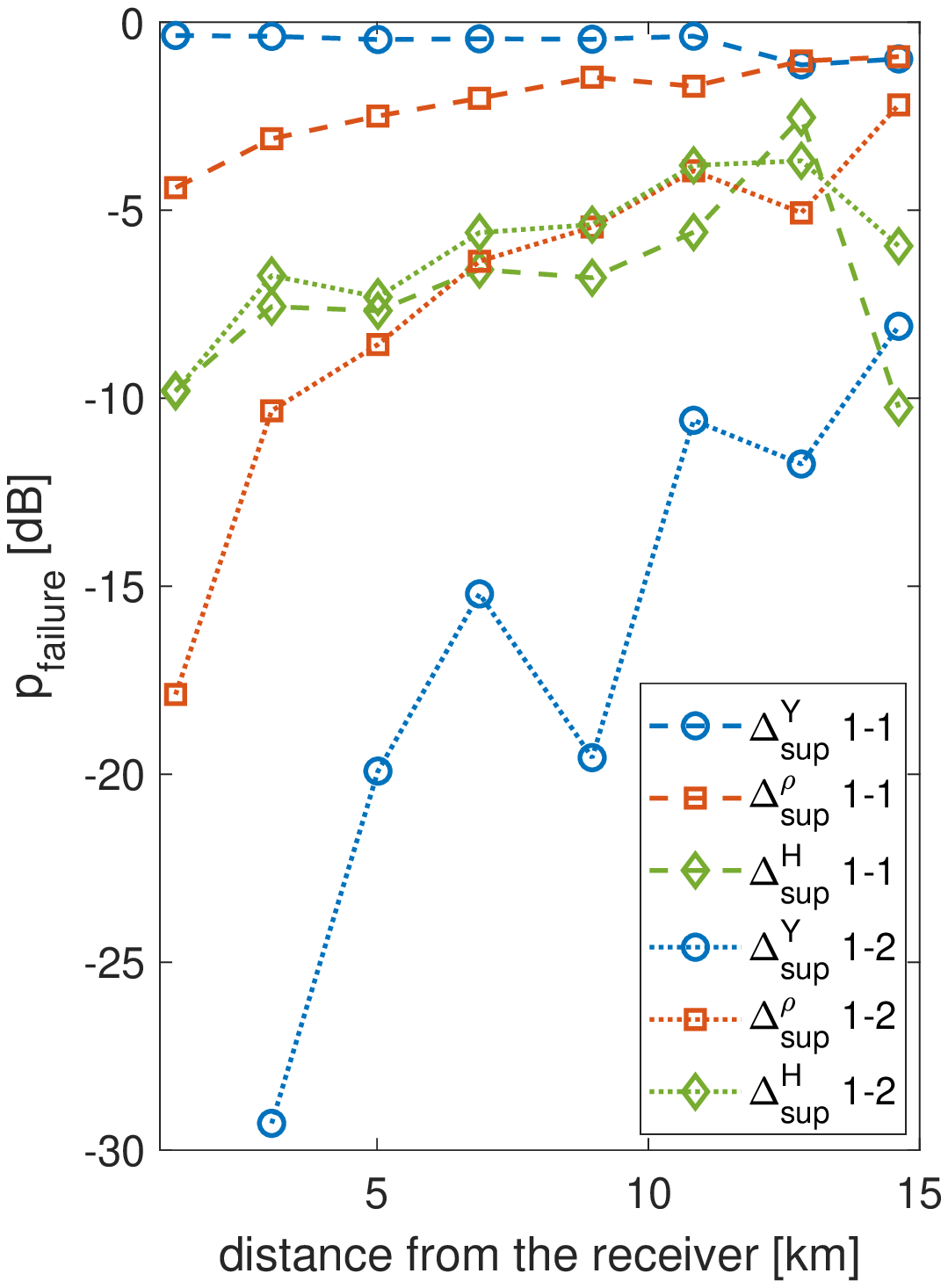}
\par\end{centering}

}\caption{$p_{failure}$ when measuring $\mathbf{Y}_{\mathbf{in}}$, $\bm{\rho}_{\mathbf{in}}$
and $\mathbf{H}_{\mathbf{tot}}$. a) Comparison of the superposition
and chain models, b) comparison of SISO and MIMO sensing. \label{fig:-when-measuring}}

\par\end{centering}

\end{figure}

As explained in Section \ref{sec:Anomaly-location}, there are different
ways to detect the presence of an anomaly with PLMs. It is possible
to estimate $\mathbf{\tilde{Y}_{in}}$, $\tilde{\bm{\rho}}_{\mathbf{in}}$
or $\mathbf{\tilde{H}_{tot}}$, using either the superposition or
the chain models, which lead to the computation of $\left|\Delta_{sup}\right|$
or $\left|\Delta_{ch}\right|$ respectively. We simulated the presence
of a fault in 2000 random networks and computed in each case the noise
distributions of $\left|\Delta_{sup}\right|$and $\left|\Delta_{ch}\right|$
for the three considered physical quantities, both in the presence
and absence of the anomaly.  By integrating over the overlapping
areas of the distributions, we computed the probability $p_{failure}$
of not detecting the anomaly and, vice-versa, of detecting a normal
measurement as anomalous. We remark that the values of $p_{failure}$
are not important per-se, since they depend on multiple factors. Herein,
we focus on the relation of the values of $p_{failure}$ obtained
with different methods.

The results of Fig. \ref{fig:-when-measuring} show $p_{failure}$
as function of $\hat{d}_{a}$ in all the aforementioned cases. Fig.
\ref{fig:-when-a} shows that the lowest values of $p_{failure}$
are reached when estimating $\mathbf{\tilde{H}_{tot}}$, followed
by $\tilde{\bm{\rho}}_{\mathbf{in}}$ and then $\mathbf{\tilde{Y}_{in}}$,
independently of the model used. This is related to the fact that
the presence of anomalies yields a greater variation in $\mathbf{\tilde{H}_{tot}}$
than in $\mathbf{\tilde{Y}_{in}}$, while $\tilde{\bm{\rho}}_{\mathbf{in}}$is
statistically more scattered (cfr. \cite[Fig. 7]{SGSI}). Regarding
the reliability of the models, Fig. \ref{fig:-when-a} shows that
the difference in $p_{failure}$ when using $\left|\Delta_{sup}\right|$
or $\left|\Delta_{ch}\right|$is not very pronounced. However, the
chain model constantly yields slightly better results than the superposition
model.

Fig. \ref{fig:-when-b} shows the performance increment obtained when
using MIMO instead of SISO measurements. We consider a fault placed
between a couple of conductors in a three-wire network. The $1-1$
symbol stands for a signal that has been sent on two non-faulted conductors
and received on the same pair. The $1-2$ symbol stands for a signal
that has been injected on the pair of conductors interested by the
fault and received on the other. The figure shows that, if the anomaly
is detected using a SISO modem placed between the non-faulted conductors,
$p_{failure}$ is greater than analyzing the cross-coupling transfer
function with a MIMO modem. This is particularly true when considering
admittance measurements, while almost no performance increment is
obtained when estimating $\mathbf{\tilde{H}_{tot}}$. This result
highlights the importance of using MIMO PLC modems when sensing power
line networks.

\subsection{Performance of the proposed algorithms}

\begin{figure}[tb]
\begin{centering}
\subfloat[Noise fixed (see Table \ref{tab:Parameters-used-for})\label{fig:Noise-fixed-(see}]{\begin{centering}
\includegraphics[width=0.43\columnwidth]{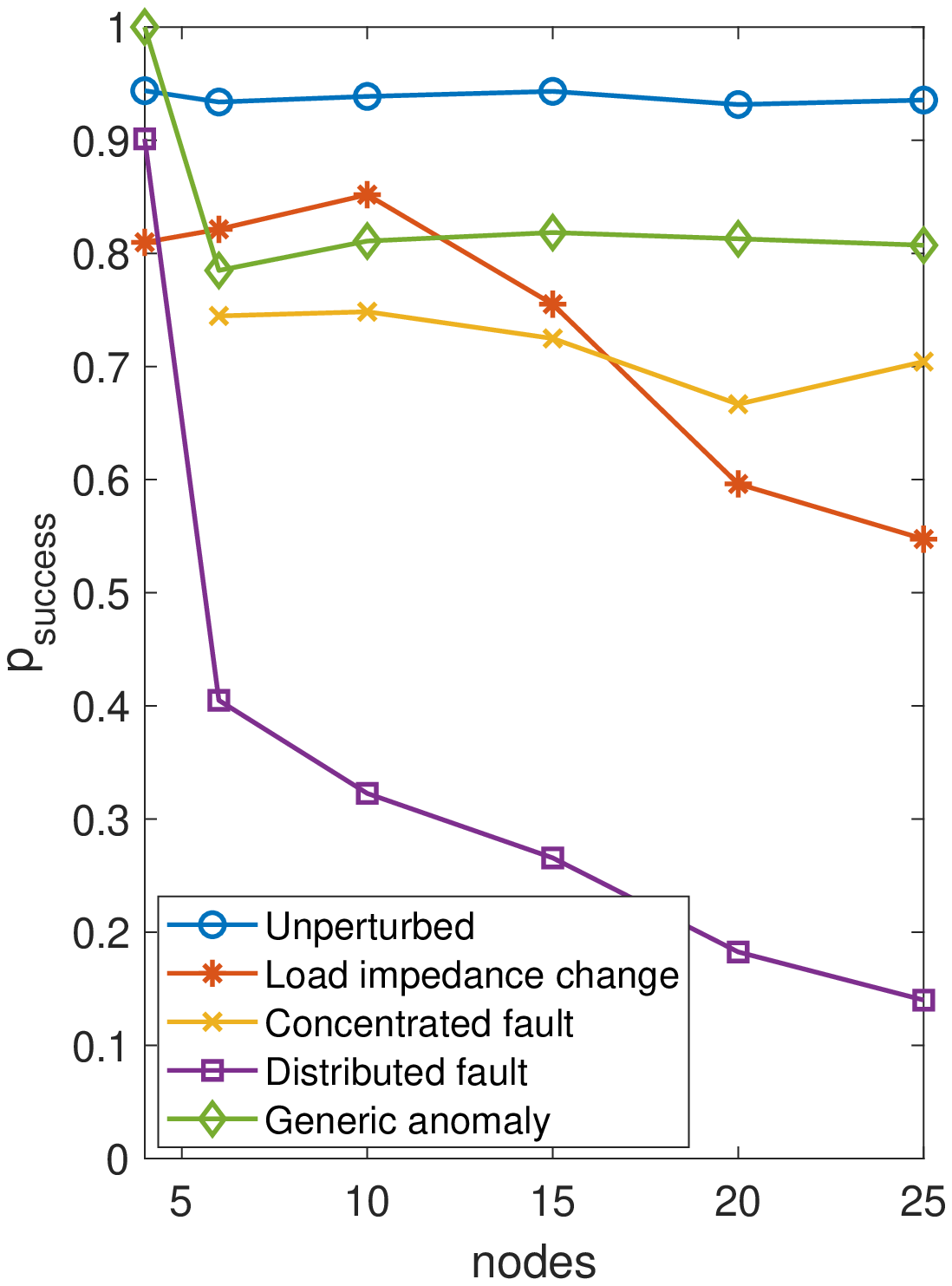}
\par\end{centering}

}\subfloat[10 nodes\label{fig:10-nodes}]{\begin{centering}
\includegraphics[width=0.43\columnwidth]{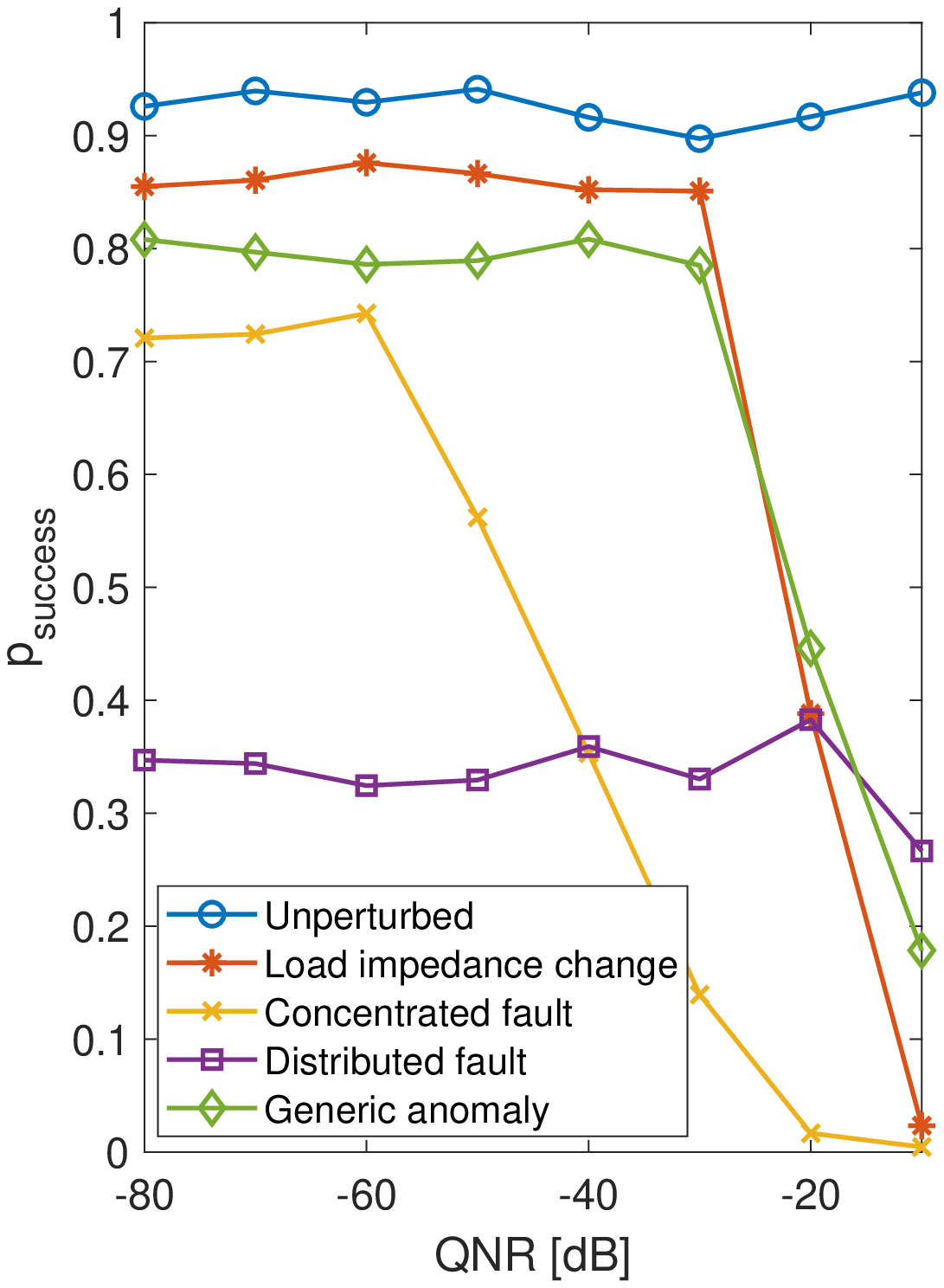}
\par\end{centering}

}\caption{Probability of correctly detecting and identifying an anomaly.\label{fig:Probability-of-correctly}}

\par\end{centering}

\end{figure}
\begin{figure}[tb]
\begin{centering}
\subfloat[Noise fixed (see Table \ref{tab:Parameters-used-for})\label{fig:Noise-fixed-(see-1}]{\centering{}\includegraphics[width=0.43\columnwidth]{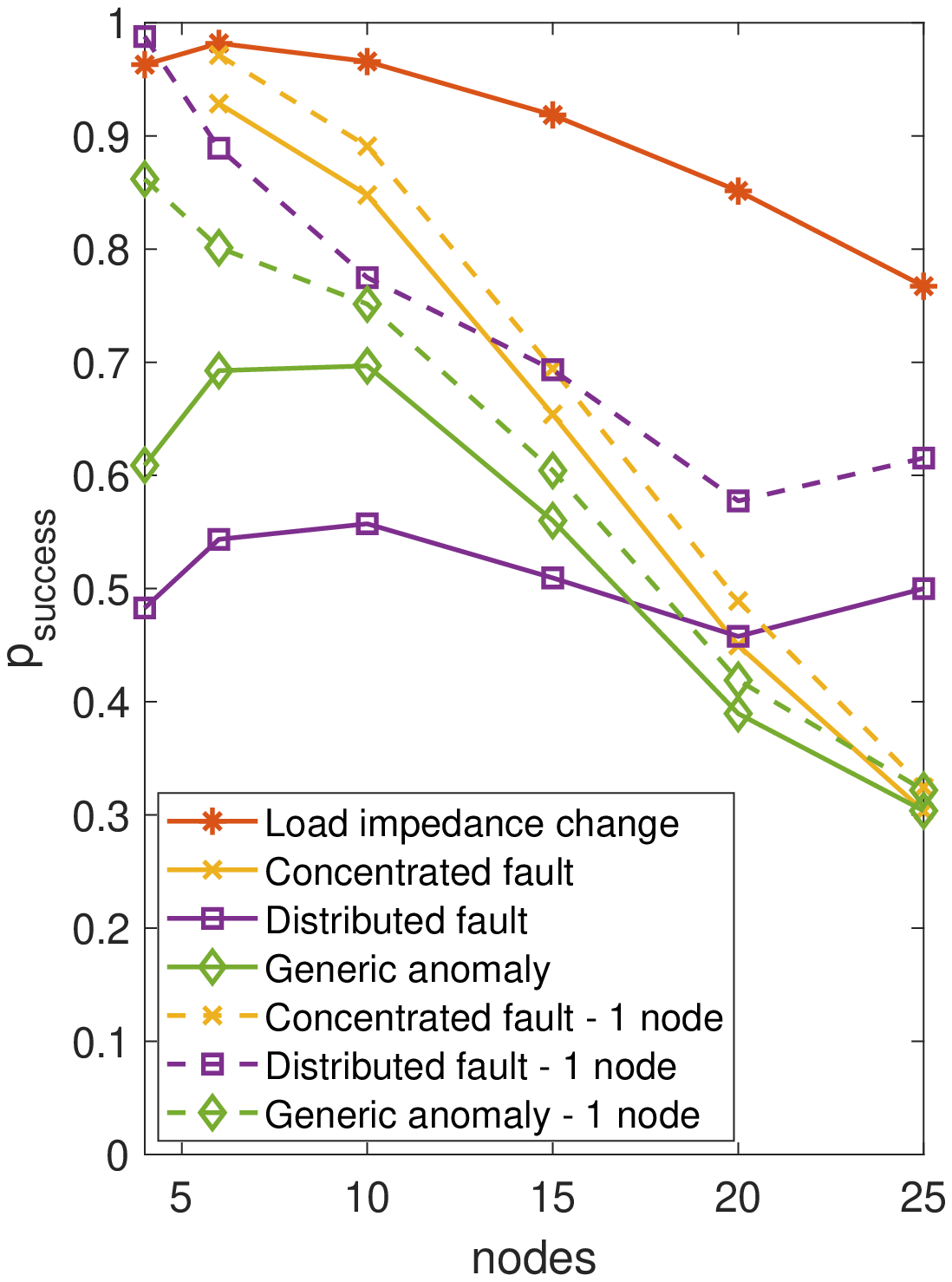}}\subfloat[10 nodes\label{fig:10-nodes-1}]{\centering{}\includegraphics[width=0.43\columnwidth]{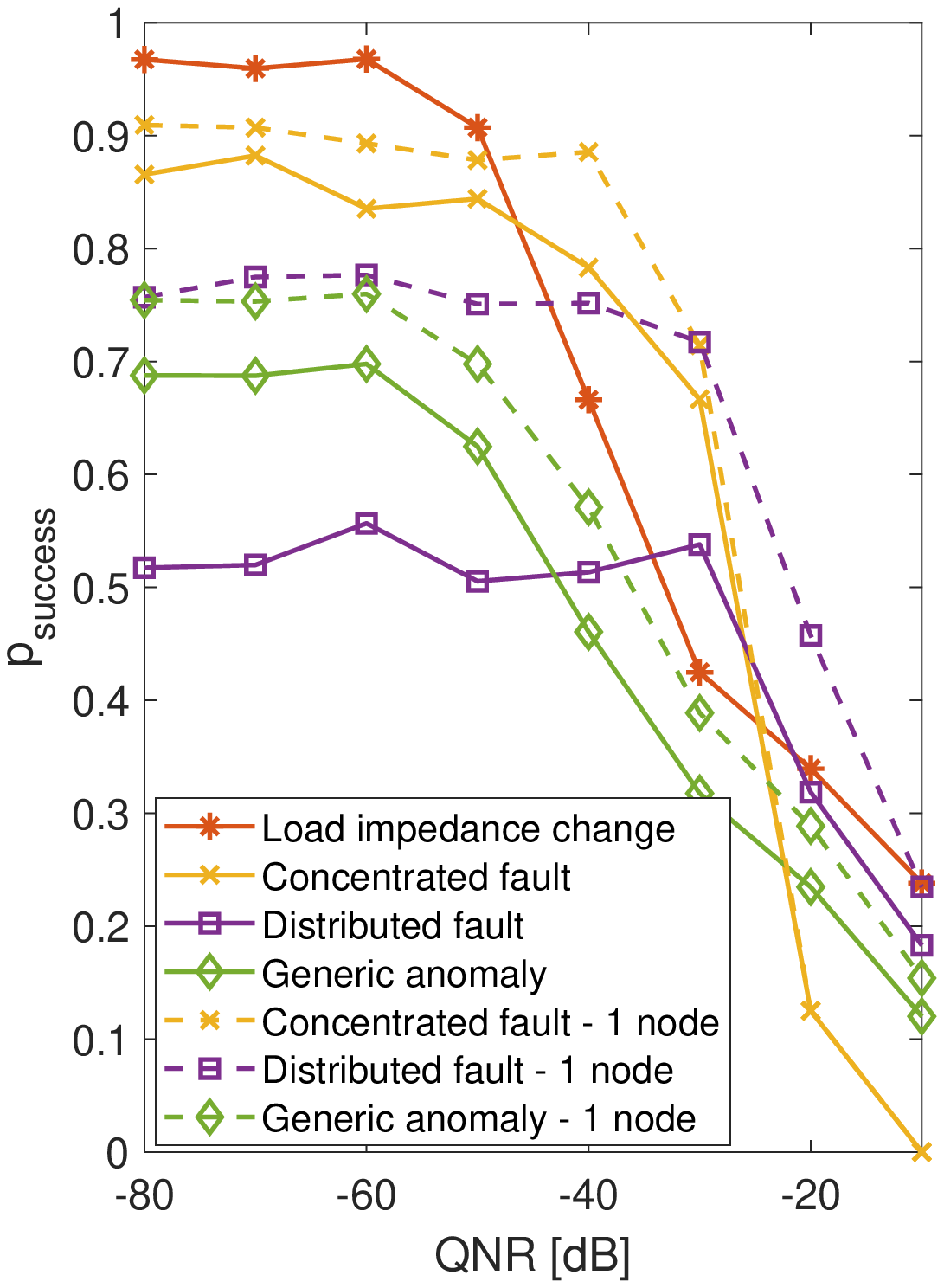}}
\par\end{centering}

\caption{Probability of correctly locating a detected anomaly.\label{fig:Probability-of-correctly-1}}

\end{figure}
In this section, we evaluate the performance of Algorithm \ref{alg:Anomaly-detection-and}
and \ref{alg:Anomaly-location-algorithm}, focusing on the estimation
of $\mathbf{\tilde{Y}_{in}}$. To this purpose, we simulate the presence
of different kind of anomalies in networks of different sizes and
compute the probability $p_{success}$ of correctly detecting, classifying
or locating an anomaly. Moreover, in order to simulate the effect
of a generic estimation algorithm, we also run a simulation where
the noise parameters of Table \ref{tab:Parameters-used-for} are no
longer used. Instead, we directly set the estimation error by modifying
the QNR, defined as \cite{7994720}
\begin{equation}
QNR=\frac{\left|X_{0}\right|^{2}}{E\left[\left|X_{N}\right|^{2}\right]},
\end{equation}
where $E\left[\cdotp\right]$ is the expectation operator, $X=X_{0}+X_{N}$
can be either $Y_{in}$, $\rho_{in}$, or $H_{tot}$, and the subscripts
$_{0}$ and $_{N}$ stand for the mean value and the noisy component
of the estimated quantity. 

Fig. \ref{fig:Probability-of-correctly} shows the performance of
Algorithm \ref{alg:Anomaly-detection-and} for varying network size
and QNR, respectively. We considered the probability of correctly
detecting and classifying the unperturbed situation and the three
considered anomalies. We also considered the probability of detecting
a generic anomaly without classifying it. The results show that the
proposed algorithm yields high values of $p_{success}$ for every
kind of anomaly when the network is very small. With increasing size
of the network, the performance of the concentrated fault case does
not significantly change while it decreases considerably in the case
of distributed faults. Fig. \ref{fig:10-nodes} shows that the algorithm
is rather resilient to noise up to values of QNR of -30 dB, with the
exception of the concentrated fault case. Low values of QNR tend not
to increase the performance of the algorithm either. This suggests
two observations: on one side, it is not needed to implement estimation
algorithms that yield very low values of QNR; on the other side, most
of the error is due to the topological structure of the network and
how the detection algorithm copes with it. Therefore, better results
can be achieved by improving the detection and classification algorithm.

Coming to the performance of Algorithm \ref{alg:Anomaly-location-algorithm}
regarding the location of anomalies, Fig. \ref{fig:Probability-of-correctly-1}
shows the probability of correctly identifying the branch where an
anomaly has occurred, when it has been correctly identified. In this
case, both the size of the network and the QNR have a significant
impact on the results. In fact, $p_{success}$ almost linearly decreases
with the number of nodes and is resilient to noise only up to a QNR
of around -50 dBm. The distributed fault case is more flat than the
others, but this is due to the fact that the detection probability
already decreases significantly with the size of the network. In Fig.
\ref{fig:Probability-of-correctly-1} we also plotted the probability
of identifying the first node of the branch where the anomaly has
occurred (there might be more ramifications). The results are in this
case significantly better, especially for the distributed fault case.
This means that, even though the exact branch might not be identified,
the set of possible faulted branches is significantly reduced.

Finally, we remark that the results presented in this paper refer
to Algorithms \ref{alg:Anomaly-detection-and} and \ref{alg:Anomaly-location-algorithm}
applied to networks with random topologies. If the algorithms had
to be applied to specific topologies or topological classes, then
they could be better tailored to the specific situation an yield better
results.

\section{Conclusions \label{sec:Conclusions}}

In this paper, we presented a framework to deploy power line communication
modems as power grid sensors, exploiting their ability to transmit
sensing signals and to acquire them. We described the modem architectures
needed for sensing and evaluated different options to estimate $\mathbf{H}$,
$\mathbf{Y_{in}}$ and $\bm{\rho}_{\mathbf{in}}$. We proposed two
monitoring techniques, namely the symbol level sensing and mains level
sensing, that allow to monitor different types of anomalies. In this
regard, we proposed two algorithms that start from the estimated channel
response at different time instants and are able to detect, classify
and locate an anomaly. The results show that correctly identifying
and locating an anomaly does not depend much on the size of the network
or on the noise, but rather on the topology of the network and how
it is taken into account by the detection and localization algorithms.
The performance of the proposed algorithms encourages further endeavors
in the area of grid monitoring with power line communications.

\bibliographystyle{IEEEtran}
\bibliography{femtocell_biblio}

\end{document}